\documentclass[conference]{IEEEtran}
\IEEEoverridecommandlockouts
\usepackage{cite}
\usepackage{amsmath,amssymb,amsfonts}
\usepackage{algorithmic}
\usepackage{graphicx}
\usepackage{textcomp}
\usepackage{xcolor}
\usepackage{balance}

\def\BibTeX{{\rm B\kern-.05em{\sc i\kern-.025em b}\kern-.08em
    T\kern-.1667em\lower.7ex\hbox{E}\kern-.125emX}}
\begin{document}

\title{Convolutional Neural Networks vs. Deformable Image Registration For Medical Slice Interpolation}

\author{\IEEEauthorblockN{Dilip Kumar Verma}
\IEEEauthorblockA{\textit{Applied Signal \& Image Processing (ASIP) Laboratory} \\
\textit{ College of Engineering and Computing Sciences} \\
\textit{New York Institute of Technology}\\
Old Westbury, NY \\}
\and
\IEEEauthorblockN{Ahmadreza Baghaie}
\IEEEauthorblockA{\textit{Applied Signal \& Image Processing (ASIP) Laboratory} \\
\textit{College of Engineering and Computing Sciences} \\
\textit{New York Institute of Technology}\\
Old Westbury, NY \\}
}

\maketitle

\begin{abstract}
Medical image slice interpolation is an active field of research. The methods for this task can be categorized into two broad groups: intensity-based and object-based interpolation methods. While intensity-based methods are generally easier to perform and less computationally expensive, object-based methods are capable of producing more accurate results and account for deformable changes in the objects within the slices. In this paper, performance of two well-known object-based interpolation methods are analyzed and compared. Here, a deformable registration-based method specifically designed for medical applications and a learning-based method, trained for video frame interpolation, are considered. While the deformable registration-based technique is capable of accurate modeling of the changes in the shapes of the objects within slices, the learning-based method is able to produce results with similar accuracy, but with much sharper appearance in a fraction of the time. This is despite the fact that the learning-based approach is not trained on medical images and rather is trained using regular video footage. However, experiments show that the method is capable of accurate slice interpolation results.\\
\end{abstract}



\section{Introduction}
Image interpolation and super-resolution as a common research area has been widely studied in various branches of image processing. Using the available image data, image interpolation/super-resolution techniques aim to enhance the resolution of the data. Such resolution enhancement can be done in both spatial domain as well as temporal domain for image sequences. Image interpolation, more specifically slice interpolation, has found wide-spread use in biomedical applications to enhance the resolution of the data acquired using biomedical imaging modalities such as CT, MRI, etc. This practice is especially useful for visualization and analysis purposes of biomedical data. In such modalities, a sequence of 2D image are acquired from the patient. However, given the limitations of the imaging systems, usually the resolution is not symmetric in all three coordinate directions as the through-plane resolution of the acquired data is significantly lower than the in-plane resolution. Such discrepancy results in step-shaped iso-surfaces and discontinuity in structures in 3D reconstructed models. Therefore, it is necessary to develop image interpolation techniques to overcome these limitations. 

Generally, slice interpolation techniques can be divided into two groups: intensity-based interpolation, and object-based interpolation. In the first group, the interpolated slices are computed directly from the intensity information of the available data, without considering the shape information of the objects contained in the images. Nearest-neighbor, linear and cubic spline interpolation methods are examples of techniques used in this group. The simplicity of such interpolation techniques results in very low computational complexity, which makes them highly popular in visualization applications. However, for analysis purposes given that the results produced by these techniques suffer from blurring artifacts on object boundaries, they are not recommended. 

In object-based methods, the shape information of the objects contained in the images are taken into account to guide the interpolation into more accurate representations of the interpolated slices. Examples of such methods can be seen in \cite{goshtasby1992matching-1, higgins1996nonlinear-2, grevera1996shape-3,lee2000morphology-4, lee2002feature-5}. Goshtasby et al. used a gradient magnitude based approach to find the corresponding points between consecutive slices and then applied linear interpolation to compute the interpolated slices \cite{goshtasby1992matching-1}. The proposed method is proven useful when the shape difference is small as the search domain is limited to small neighborhoods. However, such assumption is not generally true in practical applications. To overcome the limitations, other techniques such as column fitting interpolation \cite{higgins1996nonlinear-2}, shape-based interpolation \cite{grevera1996shape-3}, morphology-based interpolation \cite{lee2000morphology-4}, and feature-guided interpolation \cite{lee2002feature-5} have been proposed. To find pixel-wise correspondence between pixels of images, registration and/or optical flow estimation methods can be employed. Such correspondence can be formulated by taking into account intensity or feature information of images \cite{baghaie2014fast-6, baghaie2015dense-7, baghaie2017dense-8}. A relatively new group of techniques for slice interpolation use image registration as the main building block to estimate the changes in shape of the objects contained in the available slices \cite{penney2004registration-9, frakes2008new-10, leng2013medical-11, baghaie2014curvature-12, horvath2017high-13}. In this group, deformable (non-rigid) image registration serves as means to estimate the pixel-wise correspondence between the consecutive slices. 

With the introduction of deep Convolutional Neural Networks (CNNs) the field of image processing is completely transformed. In supervised CNNs, after setting the hyperparameters of the network (such as number of layers, structure of layers, evaluation functions etc.) the network is trained by introducing sets of data and the parameters of the network are trained to reduce the amount of error between the network’s output and the ground truth. Examples of such techniques can be found in 3D slice interpolation of medical images as well. Chen et al \cite{chen2018brain-14} proposed a 3D Densely Connected Super-Resolution Network (DCSRN) for slice interpolation of medical images. Peng et al \cite{peng2019deep-15} on the other hand proposed to use a 2D CNN for interpolating anisotropic brain MRI data to enhance the lower resolution along the through-plane direction. The proposed method takes advantage of CNN-based data fusion and refinement to achieve the final results. In the work of Kudo et al \cite{kudo2019virtual-16} use of conditional Generative Adversarial Networks (GANs) is explored for super-resolution of CT images 

In this paper, we explore use of CNN-based frame interpolation techniques for slice interpolation of biomedical images. More specifically, we aim to analysis the performance of adaptive separable convolutions for video frame interpolation proposed in the work of Niklaus et al \cite{niklaus2017video-17} and compare it with the registration-based approach proposed by Leng et al \cite{leng2013medical-11}. One major challenge in using deep learning-based methods in biomedical application is the limitation in access to abundance of data required to train the networks. Given this, we aim to see whether the pre-trained video frame interpolation technique which is trained on regular video footage without any human annotation can be used for biomedical slice interpolation. 

The rest of the paper is organized as follows. In section 2, an overview of both video frame interpolation and registration-based slice interpolation methods is provided. Section 3 contains experiments using biomedical volume data and quantitative and qualitative performance comparisons are presented. Section 4 concludes the paper.

\section{Methods}

\subsection{Registration-Based Slice Interpolation}
The slice interpolation proposed in \cite{leng2013medical-11} works based on multi-resolution deformable image registration. The registration model proposed in the approach is as follows. Given two input images $I_0$ and $I_1$, they are represented as continuous functions $I_0$(X) and $I_1$(X), in which $X = (x,y)$ defined in the domain $\Omega = [0,1]^2$ using bilinear interpolation. The model aims to compute two displacement maps, $U_0$ and $U_1$ that minimize the following energy function:

\begin{equation}
\label{Energy_function}
    \begin{split}
        &E(U_0, U_1) = \\
        & \int_\Omega\frac{[I_0(U_0(X))-I_1(U_1(X))]^2}{1.0 + c[I_0(U_0(X))^2 + I_1(U_1(X))^2]} dX \\
        &+ \lambda_1 \sum_{k=0}^1 \int_\Omega[||U_{kx}(X)^2||+||U_{ky}(X)^2||] dX \\
        &+ \lambda_2 \sum_{k=0}^1 \int_\Omega[||U_{kx}(X) \times U_{ky}(X)||^2] dX
    \end{split}
\end{equation}

In this equation, $U_{kx}(X)$ and $U_{ky}(X)$ are first-order derivatives of the $U_k(X)$ with respect to $x$ and $y$, respectively. The first term in the energy function is the fidelity term which aims to minimize the difference between the two deformed input images. In this formulation, bi-directional matching is used for improved performance. Also to balance the mismatch endurance of the squared intensity difference (SSD) in low-intensity and high-intensity regions, a modified SSD is employed with $c$ as a positive constant set empirically. Because of the ill-posedness of the registration problem, the energy function needs to be regularized by introducing smoothing functions. Here the first-order regularization and the area regularization terms are used respectively. These terms’ contributions in the final energy function are weighted empirically according to given images. The registration model is minimized using a geometric flow-based method. After the derivation of the registration model, the in-between slices are computed using linear or cubic spline interpolation. The avid reader is referred to the original paper \cite{leng2013medical-11} for more details on the implementation.

\subsection {Convolutional Neural Network-Based Frame Interpolation}
The CNN-based adaptive separable convolution method proposed in \cite{niklaus2017video-17} is based on a previous work by the same authors \cite{niklaus2017video-18}, that aims to reduce the computational complexity of the former approach. As is common in many image processing applications, if the higher-dimensional kernels can be estimated in separable forms, the computational complexity of the process is reduced significantly, which leads to lower computational time for both training and inference. 

The goal of the CNN-based video frame interpolation method is to synthesize a frame in-between the two input frames $I_0$ and $I_1$. For each output pixel in the interpolated frame, a pair of 2D convolution kernels are estimated, $K_1(X)$ and $K_2(X)$, to compute the intensity value of the interpolated pixel as:
\begin{equation}\label{Interpolation}
    I_{interp}(X) = K_1(X)\otimes P_1(X)+K_2(X)P_2(X) 
\end{equation}

In this formulation, $P_1$ and $P_2$ are patches centered at location ${X = (x,y)}$ in the input frames and $\otimes$ is the convolution operator. The kernels are estimated to represent both displacement and re-sampling information for the interpolation procedure. Given the high computational complexity of estimating these 2D kernels for large displacements, the aim is to estimate a pair of 1D kernels, for both the horizontal and vertical directions, for each 2D kernel as $<k_{1v},k_{1h}>$ and $<k_{2v},k_{2h}>$ where: 

\begin{equation}
\label{Estimate 1D Kernels }
    K_1 = k_{1v} \otimes k_{1h}\ \ \text{and} \ \ K_2 = k_{2v} \otimes k_{2h}
\end{equation}

The neural network architecture consists of a contracting component (encoder) for feature extraction of the two input frames and an expanding component (decoder) to perform upsampling and dense prediction. Skip connections are incorporated in the network to connect the layers from the contracting component to the layers of the expanding component. The last expanding layer is then connected to four sub-networks, each estimating one of the four required 1D kernels. For the layers in the contracting component, stacks of $3\times3$ convolution kernels with Rectified Linear Units (ReLU) combined with average pooling are used. As for the layers in the expanding component, use of bilinear interpolation is considered. 

A combination of two loss functions is used for training of the network. The first is based on the $L_1$ norm of the intensity differences between the interpolated image and the ground truth. The second one, noted as perceptual loss, is based on the $L_2$ norm of the high-level feature differences between the interpolated image and the ground truth:
\begin{equation} \label{Interpolate_orig}
    \begin{aligned}
    L_1 &= \Big|\big| I_{interp} - I_{orig} \big|\Big|_1 & \\
    L_F &= \Big|\big| \phi(I_{interp}) - \phi(I_{orig}) \big|\Big|^2_2
    \end{aligned}
\end{equation}
where $\phi$ is the feature extraction function. It is reported in \cite{niklaus2017video-17} that the \textbf{relu4\_4} layer of the VGG-19 network \cite{simonyan2014very-19} is used for feature extraction. 

The training is done by randomly selecting 250,000 data samples, each containing $150\times  150$ patches from high-quality YouTube videos that contain sufficiently large motions. Random data augmentation is done on the fly. As for the kernel sizes, kernels of size 51 are used. For more details on the implementation and training of the network, the reader is referred to the original work \cite{niklaus2017video-17}.

\section{Results And Discussions}

To assess the performance of the two methods for slice interpolation of biomedical volume images, two sets of data are used here. For the first set, a sequence of chest CT images is considered \cite{leng2013medical-11}. The chest CT sequence consists of 69 slices with size $256\times256$. The even slices are removed and then interpolated by odd slices using the two registration and CNN based methods. The second data set (RESECT) is a series of brain MRI images consisting of 391 slices of size $290\times281$ \cite{xiao2017re-20}. Similar to the 
chest sequence, here, the even slices are removed and the odd slices are used for interpolation of the missing slices. 

Three subjective/objective metrics are used for performance assessment: Peak Signal to Noise Ratio (PSNR), Structural SIMilarity (SSIM) \cite{wang2004image-21}, and Brenner Sharpness \cite{brenner1976automated-22}. While PSNR and SSIM measure the performance with respect to the ground truth, the Brenner Sharpness belongs to the class of blind quality assessment methods since it measures the sharpness of the interpolation results independent of the ground truth. 

To compute the PSNR, first we need to compute the Mean Squared Error (MSE) between the ground truth and the interpolation result. Assuming $F$ and $\hat{F}$ as the ground truth and estimated images respectively, the MSE can be defined as:

\begin{equation}\label{MSE}
    MSE = \frac{1}{N} \sum_{i=1}^{N} (f_i - \hat{f_l})^2
\end{equation}
where $f$ and $\hat{f}$ are the $i^{th}$ pixel of the ground truth and estimated images respectively, and $N$ is the total number of pixels. Having the MSE, PSNR can be defined as:

\begin{equation}\label{PSNR}
    PSNR = 10\ log_{10} \frac{L^2}{MSE}
\end{equation}
where $L$ is the dynamic range of pixel intensities in the images.

For the SSIM, three different components play significant roles: luminance, contrast ratio and structure. The simplified equation for SSIM can be written as \cite{wang2004image-21}:

\begin{equation}\label{SSIM}
    SSIM(F,\hat{F})= \frac{(2\mu_F \mu_{\hat{F}} + C_1)(2 \sigma_{F\hat{F}} + C_2)}{(\mu^2_F+\mu^2_{\hat{F}}+C_1)(\sigma^2_F+\sigma^2_{\hat{F}}+C_2)} 
\end{equation}
where $\mu_{F}$ and $\mu_{\hat F}$ are the averages and $\sigma_F$ and $\sigma_{\hat F}$ are the variances of the ground truth and the estimated image, respectively while the $\sigma_{F \hat F}$ is the covariance value. $C_1$ and $C_2$ are constants defined as $C_1=(0.01 \times L)^2$ and $C_2=(0.03 \times L)^2$. 

For the Brenner Sharpness measure, the squared sum of the image’s first derivatives in both horizontal and vertical directions is calculated \cite{brenner1976automated-22}.

Fig. 1 shows the average performance comparison of slice interpolation for the two registration-based and CNN-based interpolation methods for the chest CT sequence. From left to right, PSNR, SSIM and Brenner Sharpness are depicted, respectively. For this dataset, the CNN-based method’s average performance metrics are 26.45, 0.9088, 1940 while the performance metrics for the registration-based method are 26.85, 0.9090, 1500, for PSNR, SSIM and Brenner Sharpness, respectively. While the performance of the CNN-based method is inferior to that of registration-based interpolation in terms of PSNR, in terms of SSIM the two perform similarly, and the CNN-based method produces much sharper results as is evident from the Brenner Sharpness measure. The inferior performance in PSNR can be attributed to the fact that the CNN-based method used here relies on weight matrices that are trained on regular video datasets and not on medical datasets.

Fig. 2 provides a qualitative comparison of the performance of the two methods for the chest CT sequence. In the top row, the ground truth, the result of registration-based slice interpolation, and the result of CNN-based slice interpolation are shown respectively. In the bottom row, the difference between the two surrounding slices that are used for the interpolation, as well as the difference between the results of registration-based and CNN-based methods with respect to the ground truth are shown. Close inspection of the results reveal that the two methods perform almost identically in terms of their differences with the ground truth data. The result of the registration-based method suffers from smoothing while the result of the CNN-based method is much sharper.

\begin{figure}[!htb]
  \begin{center}
  \includegraphics[width=3.5in]{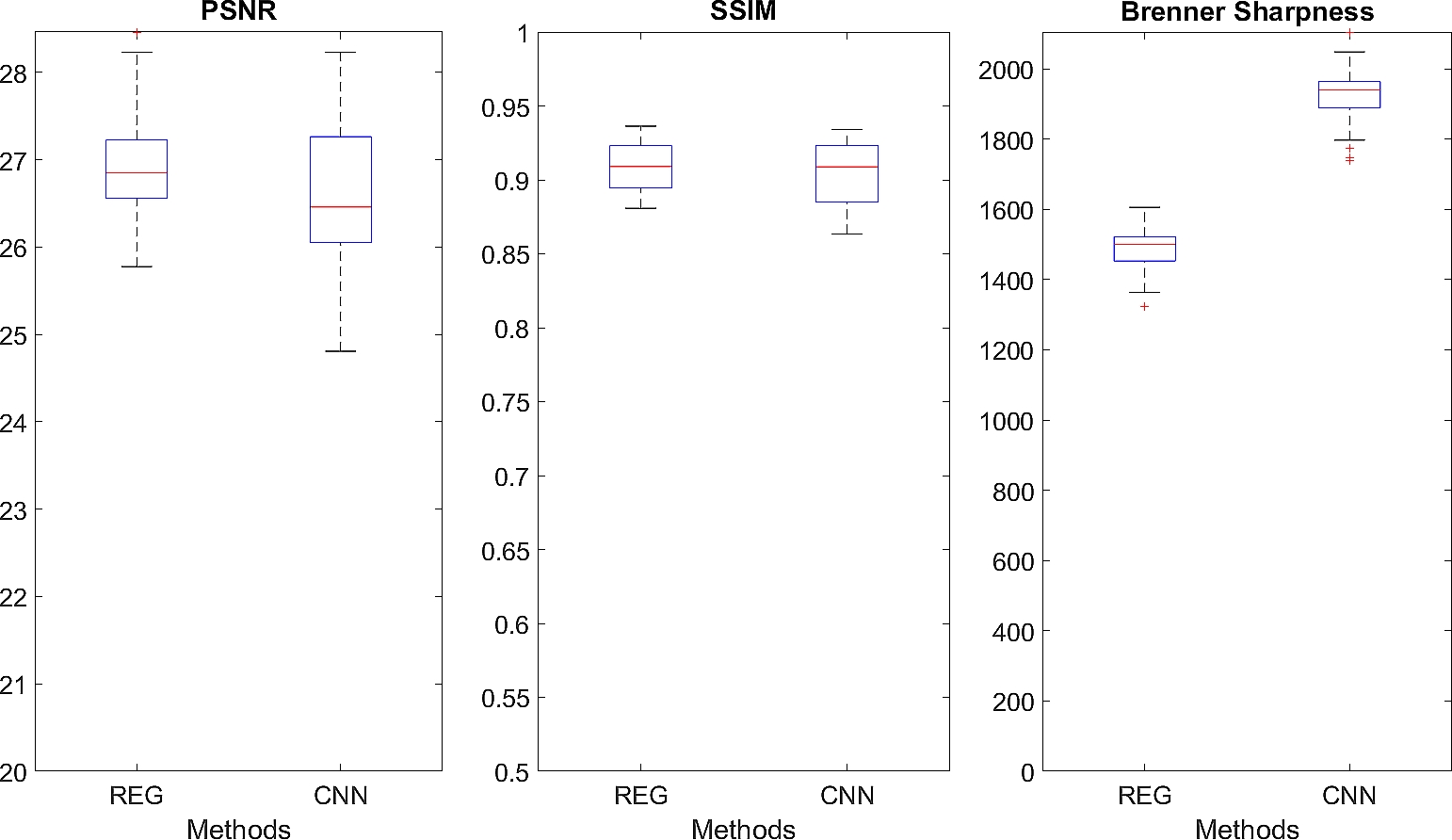}
  \caption{Average performance comparison of the registration-based (REG) and the learning-based (CNN) for the chest CT images using PSNR, SSIM and Brenner Sharpness as performance metrics.}
  \end{center}
\end{figure}

\begin{figure}[!htb]
  \begin{center}
  \includegraphics[width=3.5in]{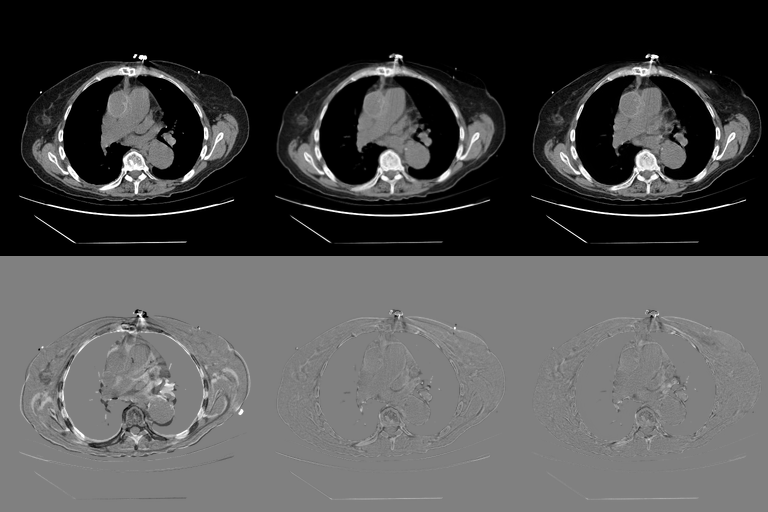}
  \caption{Sample interpolated slices from the chest CT sequence. Top row: the ground truth, the results of registration-based, and CNN-based slice interpolation, respectively. Bottom row: the difference between the two surrounding slices used for the interpolation, the difference between the results of registration-based and CNN-based with respect to the ground truth, respectively}
  \end{center}
\end{figure}

Fig. 3 shows the average performance of the two methods for the RESECT brain MRI dataset. Similar to Fig. 1, on the left the average PSNR is shown while average SSIM and average Brenner Sharpness are shown in the middle and right panels of the figure. For this dataset, the CNN-based method’s average performance metrics are 34.19, 0.9598, 1574 while the performance metrics for the registration-based method are 33.49, 0.9669, 1290, for PSNR, SSIM and Brenner Sharpness, respectively.

Fig. 4 provides a qualitative comparison of the performance of the two methods for the RESECT brain MRI dataset. As before, the ground truth, as well as the results of registration and CNN-based methods are shown in the top row, while the difference images are shown in the bottom row. Visual comparison of the results reveal that both methods perform similarly while the result of the CNN-based method is much sharper than the registration-based method.

Average computational time can also be compared. For the registration-based method the C implementation provided by the authors of the original paper are used. As for the CNN-based method, the Python implementation provided by the authors on GitHub is used. In general, given that Python is an interpreted language, its computational times are much slower than C codes. Despite this, our experiments showed that the computational time needed for slice interpolation using the CNN-based approach is much lower than that of the registration-based approach. This is to be expected, since deep learning algorithms are generally fast for inference with trained models. For the chest CT images, the computational time for interpolating 35 in-between slices of size $256\times256$ is 320 seconds and 93 seconds for the registration-based and CNN-based methods, respectively. For the RESECT data set, the computational time for interpolating 195 in-between slices of size $290\times281$ is 1965 seconds and 777 seconds for the registration-based and the CNN-based methods, respectively.

\begin{figure}[!tb]
  \begin{center}
  \includegraphics[width=3.5in]{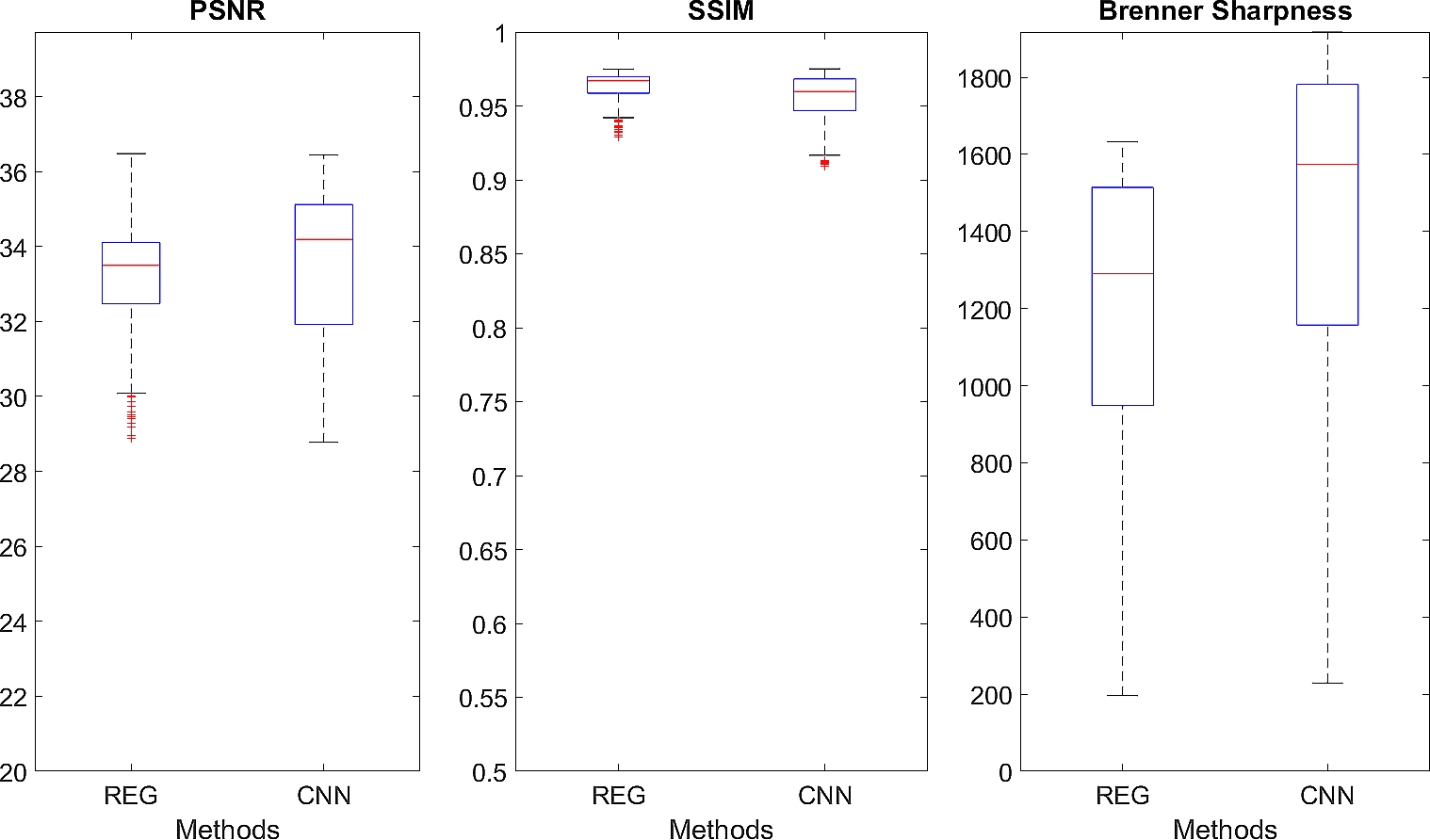}
  \caption{Average performance comparison of the registration-based (REG) and the learning-based (CNN) for the brain RESECT MRI images using PSNR, SSIM and Brenner Sharpness as performance metrics.}
  \end{center}
\end{figure}

\begin{figure}[!tb]
  \begin{center}
  \includegraphics[width=3.5in]{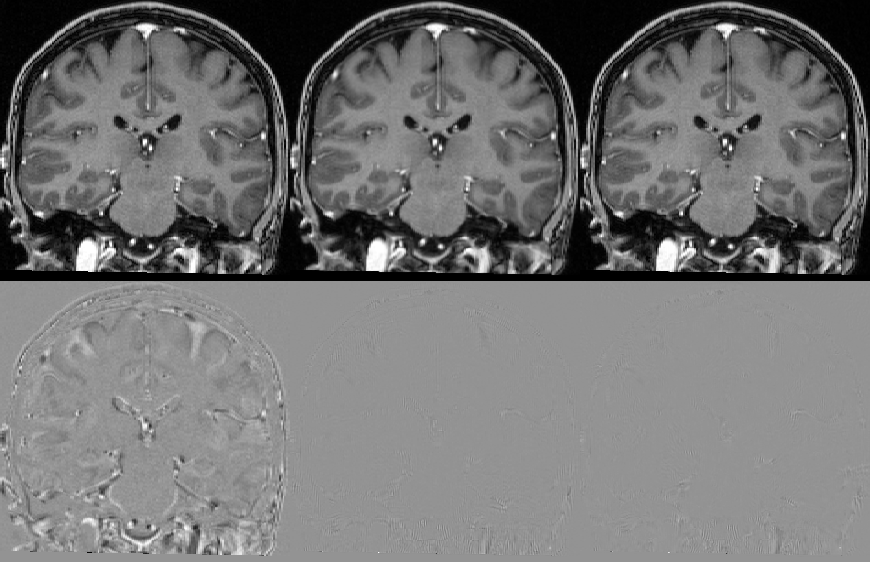}
  \caption{Sample interpolated slices from the RESECT brain MRI sequence. Top row: the ground truth, the result of registration-based, and CNN-based slice interpolation, respectively. Bottom row: the difference between the two surrounding slices used for the interpolation, the difference between the result of registration-based and CNN-based with respect to the ground truth, respectively.}
  \end{center}
\end{figure}

\section {Conclusion}

In this paper use of registration-based and learning-based methods for slice interpolation of medical images is explored. For the registration-based technique, the slice interpolation is formulated as a linear/cubic interpolation combined with a deformable registration to model the variations in the shapes of the objects contained in the available slices \cite{leng2013medical-11}. As for the learning-based approach, a deep convolutional neural network architecture is used to account for both displacement analysis and frame synthesize by taking advantage of separable convolutional kernels to reduce the computational complexity, in both training and inference steps \cite{niklaus2017video-17}. Even though the learning-based method is trained on regular video footage, and not on actual medical volume images, it is capable of producing highly accurate results in a fraction of computational time when compared with the registration-based method. This shows the great capability of the learning-based methods in such applications. Given that applicability of these techniques in medical image processing is ultimately to help in improving the processes for analysis and visualizations, it is necessary to incorporate domain knowledge into the models for a more truthful performance. This is left for future research.

\balance

\bibliographystyle{./bibliography/IEEEtran}
\bibliography{./bibliography/IEEEabrv,./bibliography/IEEEexample}


\end{document}